# Capturing social media expressions during the COVID-19 pandemic in Argentina and forecasting mental health and emotions

ANTONELA TOMMASEL
ISISTAN, CONICET-UNICEN
Tandil, Buenos Aires, Argentina
antonela.tommasel@isistan.unicen.edu.ar

ANDRES DIAZ PACE
ISISTAN, CONICET-UNICEN
Tandil, Buenos Aires, Argentina
andres.diazpace@isistan.unicen.edu.ar

JUAN MANUEL RODRIGUEZ
ISISTAN, CONICET-UNICEN
Tandil, Buenos Aires, Argentina
juanmanuel.rodriguez@isistan.unicen.edu.ar

DANIELA GODOY
ISISTAN, CONICET-UNICEN
Tandil, Buenos Aires, Argentina
daniela.godoy@isistan.unicen.edu.ar

**Purpose.** We present an approach for forecasting mental health conditions and emotions of a given population during the COVID-19 pandemic in Argentina based on language expressions used in social media. This approach permits anticipating high prevalence periods in short- to medium-term time horizons.
**Design.** Mental health conditions and emotions are captured via markers, which link social media contents with lexicons. First, we build descriptive timelines for decision makers to monitor the evolution of markers, and their correlation with crisis events. Second, we model the timelines as time series, and support their forecasting, which in turn serve to identify high prevalence points for the estimated markers.
**Findings.** Results showed that different time series forecasting strategies offer different capabilities. In the best scenario, the emergence of high prevalence periods of emotions and mental health disorders can be satisfactorily predicted with a neural network strategy, even when limited data is available in early stages of a crisis (e.g., 7 days).
**Originality.** Although there have been efforts in the literature to predict mental states of individuals, the analysis of mental health at the collective level has received scarce attention. We take a step forward by proposing a forecasting approach for analyzing the mental health of a given population (or group of individuals) at a larger scale.
**Practical implications.** We believe that this work contributes to a better understanding of how psychological processes related to crisis manifest in social media, being a valuable asset for the design, implementation and monitoring of health prevention and communication policies.

## 1. Introduction

The COVID-19 pandemic has brought changes to people's social behavior, but also posed challenges for governments, public order and mental health. For instance, the World Health Organization has expressed concerns over the mental health and psycho-social consequences of both the pandemic and its preventive policies, as they might increase loneliness, anxiety, depression, drug use, and suicidal behavior, among others (Kumar and Nayar, 2020). Given the central role that social media plays in people's life, their users have been massively expressing their thoughts and concerns regarding the pandemic. Furthermore, since social media are a bidirectional channel, the information posted can also affect these users and their mental health. This setting provides an opportunity for analyzing the pandemic effects on societal behaviors based on the social media activity that the pandemic generates, and how such activity connects with existing bodies of knowledge about mental health and emotions.

The analysis of language and sequences of textual exchanges in social media can provide rich information about individuals' behaviors, which in turn, can provide insights on how a given crisis evolves, how individuals cope with the crisis, what their needs are, and how their mental health changes, among others. The early identification of trends in mental health conditions of individuals (at the collective level) becomes crucial for decision makers when developing effective interventions or designing messaging strategies to support the affected population (Holmes et al., 2020). For instance, during a period of high prevalence of Twitter contents alluding to mental health issues (after 200 days of lock-down), the authorities decided to launch a campaign of mental health prevention both in social media and as part of diagnosing and contact tracing actions.

In this work, we present an approach for estimating mental health conditions and emotions of a given population during crises, like the COVID-19 pandemic, based on the language expressions used in social media. Mental health conditions and emotions are captured via lexical categories, referred to as *markers*, which link social media contents with well-known lexicons, such as Empath (Fast et al., 2016) and SentiSense (de Albornoz et al., 2012). Our approach works in two stages. First, we enable the construction of descriptive timelines for decision makers to monitor the *evolution of markers* and their correlation with certain crisis events or actions. As part of this stage, we detect *peaks* or *change-points* in the timelines, which represent time periods in which the observed markers undergo a substantial change with respect to previous observations. For example, high prevalence of anxiety and depression markers were observed around the time of the announcement of the first lock-down extension. Second, since the evolution of the markers can be seen as time series, we support *marker forecasting* for a given time horizon. This way, decision makers can assess what-if scenarios (in terms of future marker values and corresponding peaks), and plan for possible interventions to cope with the crisis. Our approach has been applied to a large collection of Twitter data related to the COVID-19 situation in Argentina (Tommasel et al., 2020), in which we studied mental health markers for three disorders (anxiety, depression and stress), as well as for emotions (positive and negative). Each type of disorder or emotion will be referred to as a *dimension* analysis and consists of a combination of makers. We explored both conventional time series and neural network techniques for forecasting high prevalence peaks in the series, which would suggest a potential deterioration of mental health in the population. The existence of peaks in the real series was manually validated by assessing the matching between the discovered peaks and COVID events in Argentina. Then, we analyzed the matches and coverage between those peaks and the peaks discovered based on the forecast markers and dimensions. Results showed that neural network techniques were capable of accurately identifying the periods with high prevalence of emotions and mental health disorders.

We believe that this work contributes to a better understanding of the manifestation of psychological processes related to crises as they reflect on Spanish-based social media and their users. Thus, the work has potential for informing the design of public health policies oriented to people's mental well-being during crises. Furthermore, the proposed approach is not tied to the Argentinian case-study, as it can be applied to other large-scale data streams or include other types of markers and dimensions in the analysis.

The rest of the article is organized as follows. Section 2 presents background concepts about mental health as well as some related works. Section 3 describes the approach for deriving mental health markers based on existing sociological theories, and then forecasting the prevalence of mental health markers. Section 4 describes the study of the COVID-19 dataset using the approach. Finally, Section 5 presents the conclusions and discusses future lines of work.

## 2. Background and related work

Since the beginning of the COVID-19 pandemic (as well as in previous crises), social media have become a rich information source for exposing the phenomenon, people's reactions and its effects. Besides disrupting life quality, crises often create a burden of mental health conditions by affecting individuals' expectations of the future, challenging their world view and even triggering emotional reactions (Kumar and Nayar, 2020). Hence, not surprisingly, several quantitative analyses based on social media have been conducted to detect the presence of mental health disorders (such as depression, suicidality and anxiety) and their symptomatology (Chancellor and Choudhury, 2020) through the application of natural language processing and psycho-linguistics techniques. For example, Gruebner et al. (2017) and Lin and Margolin (2014) aimed at identifying the basic emotions of Twitter users during Hurricane Sandy in 2012 and the Boston Marathon bombing in 2013, respectively. Gruebner et al. (2017) complemented the emotions analysis considering the geographic information of tweets to infer clusters of high emotion prevalence. As regards COVID-19, Li et al. (2020) and Hou et al. (2020) explored Weibo to analyze the impact of the pandemic on mental health based on the LIWC (Pennebaker et al., 2001) categories related to emotions and concerns. Li et al. (2020) compared the prevalence of the selected LIWC categories on a two-week period before and after the declaration of the outbreak. The study showed that negative emotions and sensitivity to social risks increased after the outbreak, while positive emotions decreased. Hou et al. (2020) focused on analyzing public emotion responses not only to epidemiological events, but also to the government's announcements between December 2019 and February 2020. The authors reported anxiety, sad and anger peaks after certain triggering events.

As shown by the works above, social media offer the opportunity for monitoring and analyzing stressing situations at a massive scale. Nonetheless, the analyses were descriptive and retrospective in nature, in the sense that mental health states and trends were determined by looking at collections of messages already shared by individuals. This kind of analysis limits the possibility of acting upon the crisis in the short term, for instance, by predicting possible health states based on previous observed states that can alert authorities about mental risky situations.

Along this line, several works have attempted to predict mental health disorders (such as depression, anxiety, stress) and well-being, based on individuals' physical activity information including behavioral, mobility and sleeping patterns. For example, Umematsu et al. (2019) used recurrent deep learning models based on data extracted from wearable sensors, mobile phones, and behavioral surveys from a week to predict the stress level (self-reported by the individuals) of the next day. Also based on deep learning, Suhara et al. (2017) aimed at forecasting depression based on individuals' self-reports of activities and moods at different times of the day. The determination of whether an individual was depressed depended on an assessment of the responses on a given day. Similarly, Reece et al. (2017) aimed at predicting whether an individual would be diagnosed with depression based on analyzing the presence of LIWC categories in social media posts. Predictions were based on a Hidden Markov Model and showed that three months prior to diagnosis, depressed subjects showed a marked rise in the probability of being in a depressed state, which decreased three months after the actual diagnosis. Both works focused on predicting depression at an individual level, instead of at a collective (or societal) level, as we proposed in this work.

Although there have been efforts in the literature that studied the manifestation of mental health issues in social media, and some of them have tried to predict mental states of individuals, the analysis of mental health at the collective level has received less attention. We take a step forward in this direction by proposing a forecasting approach for collective marker trends, based on the study of a large dataset of COVID-19 tweets.

## 3. Materials and methods

The proposed approach involves mainly two stages: i) data preparation and operationalization of sociological theories in terms of markers, and ii) forecasting of markers and identification of their high prevalence periods. The first stage deals with data collection (e.g., tweets), pre-processing, and matching of the pre-processed data according to predetermined lexicons. The result of this stage is a series of charts with the evolution of the markers and the mental health (i.e., `anxiety`, `depression` and `stress`) and emotions (i.e., `negative` and `positive emotions`) *dimensions* over time, which is schematized in Figure 1 (left side). Each *marker* represents a lexical category linking social media contents with the selected lexicons. Then, *dimension* values are derived from the set of marker values, as indicated by the rounded boxes in Figure 1. On this basis, the second stage is responsible for predicting future marker values (within a given time window) and then identifying high prevalence points for the corresponding dimension, as shown in Figure 1 (right side). For computing the predictions, different time series techniques can be used.

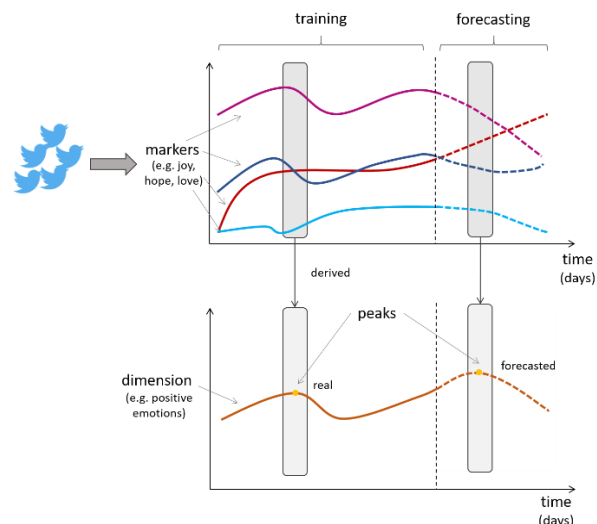

Figure 1. Evolution and forecasting of markers and dimensions

In the following sub-sections, we provide details of the approach and also the main research questions for our study.

*3.1 Data collections and pre-processing*

Our analyses are based on a large-scale sample of Twitter data collected during the COVID-19 pandemic in Argentina. The dataset, called *SpanishTweetsCOVID19*, includes more than 150 million tweets between March 1st and August 30th 2020, and is publicly available (Tommasel et al., 2020) [1]. Tweets were retrieved from the Twitter Streaming API using the Faking it![2] tool. This dataset provides a broad perspective of the mental health dynamics during the pandemic, with a center in the Argentinian situation. The selection of Twitter responds to its role as a global, real-time and mostly public mean of communication (Weller et al., 2014). We collected tweets identified as written in Spanish, and including any of a set of keywords referring to COVID (e.g., `quedateencasa`, `covid`, `covid19`, `cuarentena`, `barbijo`, `salud`, `solidaridad`, `tapatelaboca`, `mayorescuidados` or `testeomasivo`, among others) or referred to a selected user belonging to the official Argentinian offices or media (e.g., `msalnacion`, `alferdez`, `casarosada`, `infobae`, `lanacion`, `kicillofof`, `gcba` or `clarincom`, among others). We also considered tweets including geographic information located inside the Argentinian geographical bounding box. Retweets accumulated between the 15% (March) and 72% (May) monthly tweets. The number of retrieved tweets peaked in June. To keep tweets that provide enough content to evaluate aspects related to mental health, we discarded retweets, only keeping original tweets and replies. Table I summarizes the data characteristics. Hashtags appear only in the 19% of the tweets. A data inspection showed that neither politicians, media nor official government accounts included hashtags in most tweets. Collected tweets were pre-processed to remove URL, mentions and special characters. Hashtags were split into their constituent words. Spelling corrections were not applied as an inspection of randomly selected tweets showed that tweets were mostly correctly written.

Table I. Statistics of the *SpanishTweetsCOVID-19* dataset

| | | | |
|---|---|---|---|
| | Start data: 1 March 2020 | End date: 31 August 2020 | |
| Tweets | Original: 29,712, 138 (with quotes: 8,947,933) | Retweets: 106,361,088 | Replies: 15,815,975 (with quotes: 414,984) |
| | With media: 7,578,945 | With mentions: 19,494,778 | |
| | With URL: 9,917,012 | With places: 102,241 | |
| | With hashtags: 8,621,975 | number of hashtags per tweet: 46 frequency of hashtags: 13 | |
| Users | Total: 6,977,032 | | |
| | Min tweets per user: 1 | tweets per user: 4.5 ± 8.4 | Max tweets per user: 167 |

*3.2 Lexicon-based analysis for markers*

The particular choices of language that users adopt in social media for expressing their opinions and personal situations provide evidences of their state of mind. In this context, lexicons have been widely used for a text analysis of emotions, concerns, topics and health-related issues (De Choudhury et al., 2013).

In addition to emotions, this study focuses on three mental health dimensions: `anxiety`, `depression` and `stress`, which have been reported to increase due to the COVID outbreak (Kumar and Nayar, 2020). To this end, we rely on two particular lexicons: *SentiSense Affective* (de Albornoz et al., 2012) and *Empath* (Fast et al., 2016), for the psycho-linguistic analysis of emotions and mental health, respectively. The SentiSense Affective Lexicon labels 2,190 WordNet synsets according to 14 emotional categories or *markers*. Markers are divided into two dimensions: `positive` (calmness, joy, love, hope, like and anticipation) and `negative` (hate, anger, sadness, fear and disgust) `emotions`. As SentiSense is integrated with the WordNet Spanish version, it can be directly applied to the analysis of the collected tweets. On the other hand, Empath includes 200 emotional and topical human-validated categories.

We analyzed each mental health dimension based on matching the Empath categories with lexicons. Lexicons were derived from several literature sources: su Park (2012), Aldarwish and Ahmad (2017) and De Choudhury et al. (2013), as well as from the manual hand-coding of the characterizations, manifestations and symptoms of the selected disorders as defined by the National Institute of Mental Health (2020) and the Anxiety and Depression

Association of America (2020). These lexicons were semantically expanded using *FastText*[3] to capture the context in which words are used and to understand how individuals express aspects related to their mental health as in (Losada and Gamallo, 2020). At last, from these expansions, we automatically retrieved the top-10 most prevalent Empath categories (i.e., categories with the highest number of matching words with the lexicons). The selected categories are the *markers* for the associated mental health dimension. As the Empath lexicon and categories are only available in English, and given that the reliability of translating psycho-metric scales and lexicons has already been studied (Perczek et al., 2000), we automatically translated the collected tweets using the IBM Watson Language Translator [4]. The original lexicons, expansions and the translation of Empath categories can be found in the corresponding companion repository [5].

The daily prevalence of a marker was computed as the percentage of tweets on such day with at least a word matching the lexicon marker. Based on the marker prevalence in the whole time period, we model their sequence of observations and obtained their temporal distributions or *time series*. In our case, a time series models a sequence of observations about a given marker (or dimension) during a given time window. For emotions the time period spanned between March 1st - August 30th, 2020, while for mental health it did not include the last two months due to different orders of magnitude across the prevalence of markers, which hindered the analysis.

For identifying the periods of high prevalence of emotions or mental health dimensions, we search the time series for *peaks*, which can be regarded as points in which the values of the markers involved in a dimension varied altogether. These variations were presumably caused by COVID related events. For example, the first `anxiety` peak matches the days following the confirmation of the first COVID case in Argentina and the first suspension of activities. For each *marker*, we computed the gradient over the one-week smoothed time series. The smoothing reduces the impact of day-to-day variations and weekly periodicities of the time series, causes time series to respond more slowly to recent and sudden changes, thus favoring the observation of more consistent behaviors over longer periods in time. Then, we averaged the gradients for all markers to obtain the overall gradient of the *dimension*, and compute the peaks. We only kept the peaks whose prominence values were higher than the 80 percentile.

### 3.3 Forecasting markers and prevalence points

A key characteristic of a time series is the serial correlation between adjacent observations. Thus, historical values can reveal the changing trend, and this pattern can be projected to future values of the series. In our case, based on the history of mental health and emotion markers, we want to forecast the future values of such markers, which are then combined to derive dimension values. Based on the dimension values (both actual and predicted), we can anticipate the emergence of high prevalence periods. For example, departing from the `anxiety` markers during the first two weeks of March, we can forecast the markers for the following week and then compute the corresponding prevalence points for that dimension. To do so, three strategies, in increasing order of complexity, were studied: i) univariate, ii) multivariate, and iii) deep learning. The time period for observing mental health symptoms depends on the disorder under analysis. While anxiety and stress can manifest in the short-term, psychological questionnaires (such as PHQ-9 for depression) and guidelines (National Institute of Mental Health, 2020) state that symptoms can be present for at least two weeks before getting a diagnostic. Thus, for adequately capturing the mental health status, we varied the training period (for the time series strategies) between 7, 14 or 21 days. As time series forecasting often does not require long data sequences, we considered a time horizon of 7 days. This period is also useful to support decision makers in assessing preventive or counteracting measures for mental health issues.

*Univariate forecasting*
This is the simplest scenario, in which marker predictions for each time series are made separately, disregarding possible dependencies between the series. That is, markers are assumed to not affect the future value of the other markers. We used ARIMA (Box and Jenkins, 1990) and Prophet (Taylor and Letham, 2018) for this purpose. ARIMA is a classical model that blends auto-regression (i.e., a predefined number of prior observations or lag order) and moving averages in stationary series. Thus, for our marker timelines, we previously checked whether the series were stationary, and if not, we de-trended them. Prophet is an additive regression model originally developed by Facebook that works best with series with seasonal effects and a large number of historical data. It is typically robust to shifts in the trend and handles outliers well. Unlike ARIMA, Prophet does not require transformations nor pre-processing to be applied to the data.

*Multivariate forecasting*

Mental health states might have interactions among them, which are related to the notion of synchronization or response coherence (Kuppens and Verduyn, 2017). These interactions, therefore, can be present in our marker time series, but cannot be captured with a univariate analysis. Multivariate analysis, in turn, aims at modeling the dynamic relationships among the variables of a group of time series. A natural extension to univariate auto-regressive models is Vector Auto Regression (VAR) (Ltkepohl, 2007). VAR is a stochastic process that represents a group of time series as a linear function of their own past values (lag order) and the past values of all the other series in the group. This type of model has already been used in the forecast of emotions (Bringmann et al., 2018). As in ARIMA, the series are required to be mostly stationary and without seasonal trends. In addition, the group of time series should exhibit some correlation for VAR to work well. A relevant property is the forecasting ability of a variable with respect to other variables. A Granger causality test (Bringmann et al., 2018) of our time series revealed indeed interactions among most markers for the analyzed dimensions.

*Deep learning forecasting*

This is a multivariate strategy based on Recurrent Neural Networks (RNN) (Williams and Zipser, 1989), which are suited for learning problems in which data has a sequential nature, as time series analysis. RNN can also capture complex dependencies among the variables. In RNN, the connections between the neurons can form cycles, building an internal memory that facilitates learning from naturally sequential data, in which new predictions depend on the previous ones. Predictions were based on a Gated Recurrent Unit (GRU) (Cho et al., 2014) network, a particular type of RNN, which is effective for modeling varying length sequences and capturing short- to medium-range dependencies. GRUs have successfully been used for multivariate time series forecasting. They are less complex than other RNN, which translates into fewer parameters, and faster and better learning opportunities with limited data. The GRU was set to optimize the Mean Square Error loss with a RMSProp optimizer. Given that not every marker was on the same scale, data was normalized based on the quartile distribution. Like Prophet, no additional transformations of the data were required.

*3.4 Research questions*

Based on the approach above, the goal of our study was to assess whether the time series forecasts were accurate enough, particularly with respect to high prevalence periods in the dimension analysis. We departed from the assumption that the temporal evolution of the markers reflected people's emotions and mental health, and it was aligned with key pandemic-related events in Argentina. This assumption was empirically validated by the authors for the whole time period of the dataset. In the experimental work, we addressed three research questions:

- **RQ1** - How does the quality of the forecast depend on the training period?
- **RQ2** - Does the forecast performance for mental health and emotional dimensions differ?
- **RQ3** - Which time series strategy does it provide the best forecast?

**4. Data analysis and findings**

For assessing the three strategies, we used a sliding window that consisted of a training period and a forecasting horizon, and rolled out the window over the whole timeframe of the series. We varied the training period between 7, 14 or 21 days, and kept a forecasting horizon of 7 days. As the experiments were not conducted in real time, we were able to evaluate the forecasting performance by comparing the predicted values against the real ones. Three data analyses were performed for evaluating the results with respect to the research questions. First, we looked at the forecasting errors of the four strategies (ARIMA, Prophet, VAR and RNN) in terms of the Mean Absolute Percentage Error (MAPE), in order to determine differences among the strategies. The MAPE values for markers were compared using a paired statistical test, such as Wilcoxon or t-test, depending on the normality of data. The p-value was set to 0.01 and the null and alternative hypothesis were defined. The null hypothesis stated that no difference existed between the tested alternatives (for a given training period), whereas the alternative one stated that the observed differences were significant and not due to chance. Cohen's d was used to quantify the effect size between the compared values. This comparison helped us to determine the optimal number of training days per strategy. For such optimal numbers, Table II summarizes the average forecasting errors for each dimension and their markers.

Table II. Summary of Mean Absolute Percentage Errors (MAPE) for the best performing forecastings

(a) Anxiety

|  | Anger | Confusion | Disappointment | Fear | Health | Horror | Nervousness | Sadness | Shame | Suffering |
|---|---|---|---|---|---|---|---|---|---|---|
| ARIMA | 10.59 ±13.45 | 18.23 ±19.99 | 12.97 ±16.67 | 10.32 ±10.24 | 30.35 ±33.17 | 10.65 ±10.25 | 15.16 ±19.68 | 11.64 ±8.66 | 13.87 ±11.31 | 9.15 ±8.56 |
| Prophet | 11.91 ±18.79 | 15.89 ±23.15 | 15.01 ±16.16 | 13.8 ±16.74 | 34.1 ±42.47 | 18.89 ±32.88 | 19.27 ±32.63 | 13.66 ±9.52 | 13.83 ±12.27 | 9.69 ±11.33 |
| VAR | 8.88 ± 7.01 | 14.59 ± 16.13 | 13 ± 14.41 | 7.64 ± 5.91 | 15.23 ± 16.57 | 12.15 ± 13.66 | 7.79 ± 5.77 | 8.88 ± 7.11 | 14.99 ± 16.38 | 7.74 ± 6.23 |
| Deep Learning | 4.91 ± 4.9 | 6.69 ± 6.9 | 6.33 ± 7.85 | 7.2 ± 9.2 | 15.92 ± 14.45 | 8.41 ± 12.66 | 8.52 ± 10.69 | 5.91 ± 5.77 | 7.32 ± 7.48 | 5.83 ± 6.27 |

(b) Depression

|  | Disappointment | Disgust | Emotional | Neglect | Nervousness | Pain | Sadness | Shame | Suffering | Torment |
|---|---|---|---|---|---|---|---|---|---|---|
| ARIMA | 12.97 ±16.67 | 11 ±12.8 | 19.25 ±46.52 | 11.97 ±12.15 | 15.16 ±19.68 | 9.26 ±11.55 | 11.64 ±8.66 | 13.87 ±11.31 | 9.15 ±8.56 | 16.69 ±17.94 |
| Prophet | 15.01 ±16.16 | 14.07 ±19.61 | 23.62 ±45.83 | 13.61 ±14.86 | 19.27 ±32.63 | 11.53 ±17.66 | 13.66 ±9.52 | 13.83 ±12.27 | 9.69 ±11.33 | 19.52 ±21.51 |
| VAR | 9.55 ± 9.8 | 11.94 ± 11.6 | 10.37 ± 9.39 | 11.05 ± 12.55 | 8.96 ± 8.54 | 9.73 ± 9.62 | 9.07 ± 11.79 | 10.32 ± 12.88 | 9.47 ± 11.08 | 10.55 ± 9.71 |
| Deep Learning | 8.17 ± 10.37 | 6.29 ± 7.83 | 6.54 ± 7.94 | 7.19 ± 7.74 | 7.36 ± 8.51 | 6.02 ± 8.37 | 4.85 ± 4.24 | 8.16 ± 9.62 | 4.87 ± 6.19 | 11.55 ± 17.02 |

(c) Stress

|  | Anger | Disgust | Fear | Health | Neglect | Nervousness | Sadness | Shame | Suffering | Torment |
|---|---|---|---|---|---|---|---|---|---|---|
| ARIMA | 10.59 ±13.45 | 11 ±12.8 | 10.32 ±10.24 | 30.35 ±33.17 | 11.97 ±12.15 | 15.16 ±19.68 | 11.64 ±8.66 | 13.87 ±11.31 | 9.15 ±8.56 | 16.69 ±17.94 |
| Prophet | 11.91 ±18.79 | 14.07 ±19.61 | 13.8 ±16.74 | 34.1 ±42.47 | 13.61 ±14.86 | 19.27 ±32.63 | 13.66 ±9.52 | 13.83 ±12.27 | 9.69 ±11.33 | 19.52 ±21.51 |
| VAR | 25.14 ± 104.95 | 40.13 ± 81.51 | 37.59 ± 238.84 | 41.2 ± 86.01 | 59.19 ± 254.64 | 26.87 ± 143.85 | 312.24 ± 3146.58 | 24.61 ± 77.53 | 25.38 ± 75.8 | 19.48 ± 40.13 |
| Deep Learning | 4.68 ± 4.48 | 7.51 ± 8.33 | 8.03 ± 10.69 | 15.92 ± 13.38 | 7.73 ± 7.17 | 9.55 ± 11.62 | 5.48 ± 5.12 | 8.65 ± 8.6 | 5.03 ± 5.33 | 10.34 ± 16.85 |

(d) Positive Emotions

|  | Anticipation | Calmness | Hope | Joy | Like | Surprise |
|---|---|---|---|---|---|---|
| ARIMA | 2.94 ± 3.82 | 7.09 ± 7.75 | 3.47 ± 4.37 | 4.76 ± 5.49 | 2.32 ± 3.34 | 4.26 ± 4.86 |
| Prophet | 3.48 ± 4.64 | 7.35 ± 7.39 | 4.46 ± 4.82 | 5.72 ± 6.48 | 3.43 ± 4.43 | 6.05 ± 6.03 |
| VAR | 2.98 ± 4.38 | 4.9 ± 5.8 | 3.63 ± 5.06 | 4.46 ± 5.88 | 2.96 ± 4.35 | 3.09 ± 4.45 |
| Deep Learning | 2.3 ± 3.44 | 6.85 ± 7.65 | 3.05 ± 3.84 | 5.04 ± 5.66 | 2.22 ± 2.75 | 4.07 ± 4.62 |

(e) Negative Emotions

|  | Anger | Disgust | Fear | Hate | Sadness |
|---|---|---|---|---|---|
| ARIMA | 7.3 ± 8.07 | 2.54 ± 3.25 | 5.97 ± 6.85 | 14.03 ± 18.62 | 7.86 ± 7.2 |
| Prophet | 8.42 ± 7.98 | 3.17 ± 4.17 | 6.32 ± 6.55 | 13.24 ± 14.82 | 7.37 ± 6.78 |
| VAR | 6.93 ± 8.44 | 4.47 ± 5.54 | 3.95 ± 4.95 | 4.48 ± 5.85 | 8.51 ± 8.92 |
| Deep Learning | 8.01 ± 0 | 2.25 ± 2.84 | 6.3 ± 6.35 | 10.17 ± 16.25 | 7.48 ± 6.39 |

As a second analysis step, we compared the real and forecast values for each mental health or emotion marker, also based on paired statistical tests. Ideally, the forecast and real values should not differ much from each other to allow a satisfactory prediction of high prevalence peaks. At last, in the third analysis step, we compared for each dimension the peaks that resulted from the forecast markers and the peaks observed in the actual time series. To do so, we computed the hit-rate (also known as recall), which measures the proportion of true elements (i.e., peaks) discovered by the strategies. We considered a hit if a (predicted) peak was detected in a window of $n$ days before and after an actual peak in the dimension time series, with $n$ being 2, 3 or 7. In the case of $n = 7$, a hit was successful if it was found in the same enclosing week. Table III presents the hit rate results for the best performing forecastings (the best results are in bold), and Figure 2 shows for a 3-day window (marked by the grey areas) the actual and forecast peaks for each of the dimensions.

Table III. Hit Rates for the best performing forecastings for n= 2 - 3 – 7

|  | Anxiety | Depression | Stress | Positive Emotions | Negative Emotions |
|---|---|---|---|---|---|
| ARIMA | 42% - 58% - 64% | 21% - 50% - 57% | 50% - 56% - 60% | 44% - 52% - 63% | 43% - 57% - 65% |
| Prophet | 58% - 75% - 73% | 40% - 60% - 67% | 56% - 56% - 75% | 33% - 44% - 59% | 57% - 60% - 55% |
| VAR | 50% - 67% - 73% | 44% - 56% - 75% | 55% - 64% - **78%** | 47% - 63% - 57% | 55% - **70%** - 57% |
| Deep Learning | **92% - 92% - 92%** | **83% - 89% - 92%** | 64% - **82%** - 78% | **50% - 67% - 71%** | **61%** - 70% - **72%** |

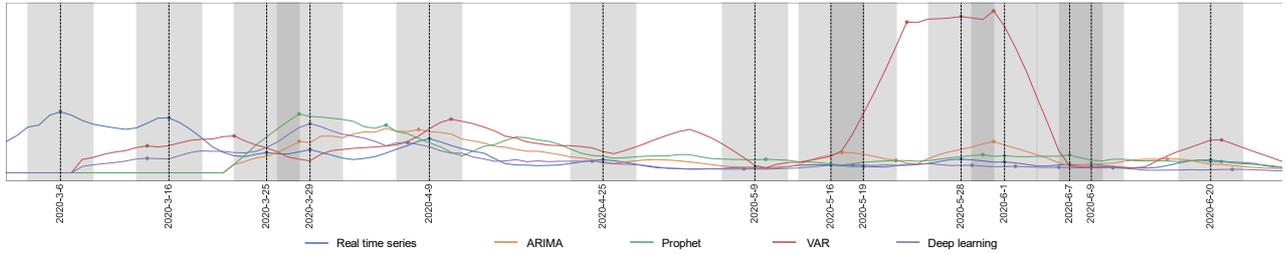

(a) Anxiety

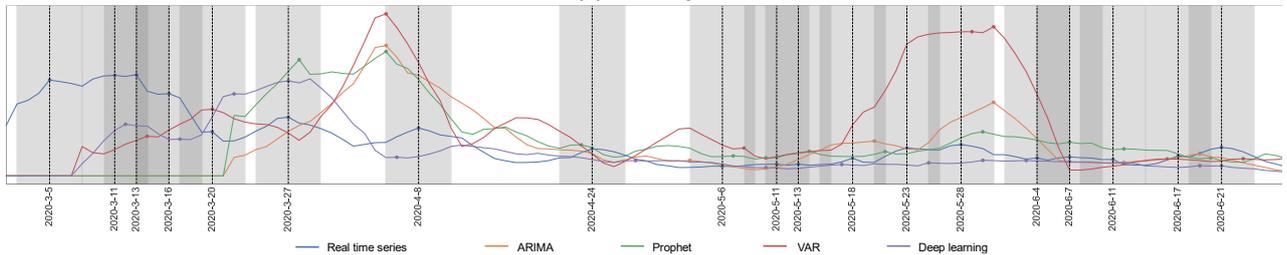

(b) Depression

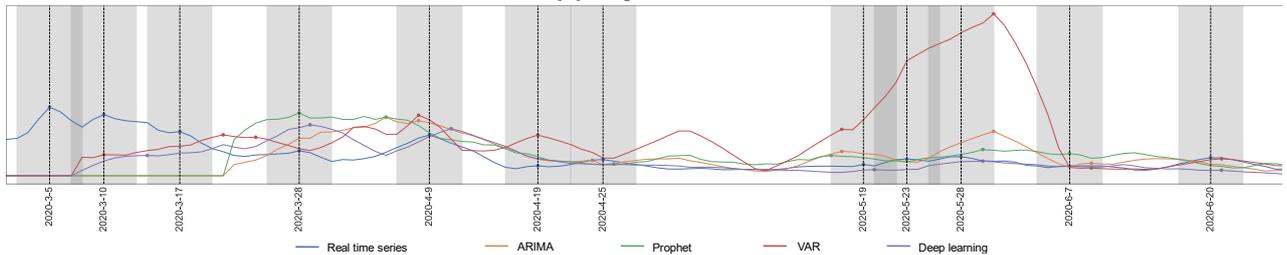

(c) Stress

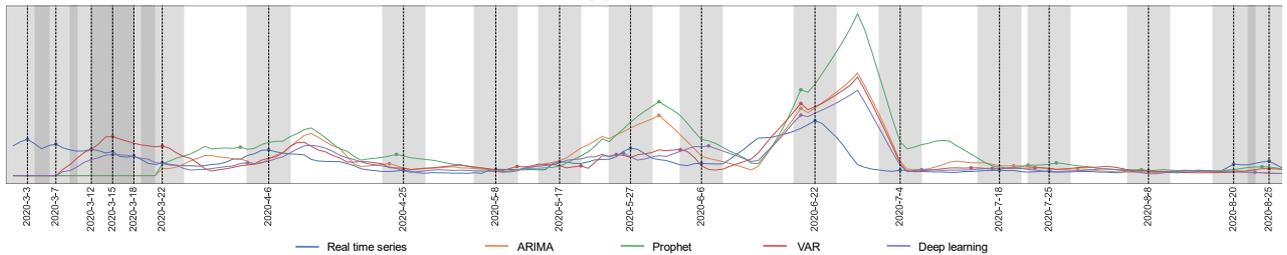

(d) Positive Emotions

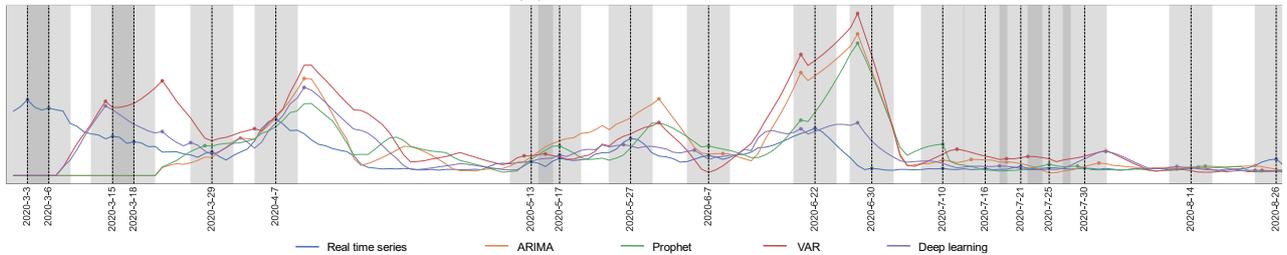

(e) Negative Emotions

Figure 2. Real and forecast prevalence and peaks for the analyzed dimensions

In the following sub-sections, we discuss the performance of each time series strategy and related findings. Extended versions of the presented tables can be found in the companion repository [5].

*4.1 Univariate forecasting*

Comparing the MAPE results for ARIMA and Prophet for the different markers and varying the training days, we observed the lowest errors when considering 21 (and even 28) training days in both cases. In most cases, these errors were statistically significantly lower than the errors for training periods of 7 or 14 days. For example, the dimension analysis for `positive emotions` showed differences for most markers, either using ARIMA or Prophet, while for the `negative emotion` dimension, differences were observed for all but the *hate* marker. Nonetheless, for some particular markers and dimensions no significant error differences were observed. For example, using ARIMA the `anxiety` dimension had no difference for the *sadness, nervousness* and *suffering* marker; while for Prophet differences were observed only for the *disappointment*. These observations disagree with those by Suhara et al. (2017), who stated that 14 days should be enough for forecasting mental health disorders, such as depression. Along this line, we argue the independent analysis (i.e., univariate forecasting) of markers is not enough for accurately forecasting them. The errors for `positive` and `negative emotions` were lower than those for the `mental health` dimensions, which implies that emotions are simpler to forecast, and that the inter-dependencies among the emotional markers are not as strong as those among the mental health markers.

When comparing ARIMA and Prophet with 21 training days, we found small-to-medium significant differences favoring ARIMA. For example, in the `emotions` dimension, significant differences were observed for every `positive emotion` but *calmness*, and for three `negative emotions` (*hate*, *anger* and *fear*). In the case of the mental health dimensions, some differences favoring ARIMA were observed for `anxiety` (*sadness*, *fear*, *horror* and *disappointment*), `depression` (*sadness*, *emotional*, *disappointment*) and `stress` (*sadness* and *fear*). In general, Prophet did not perform better than ARIMA, which could be related to the fact that Prophet needs a larger training period.

The comparison between the gradient of the real and forecast markers revealed medium-to-large statistical differences for most markers. When comparing the gradients for the real and forecast dimensions, for ARIMA differences were small and even negligible. This could imply that although the marker forecasting is not perfect regarding errors, it still captured the tendency of the dimension time series adequately (as shown in Figure 2). For Prophet, medium-to-large significant differences were observed in all cases. Regarding the hit rates in Table II, ARIMA and Prophet allowed to discover in average 40% and 49% of the peaks ($n = 2$), respectively. In average, the highest rates were observed for `negative emotions` (65% for ARIMA), and `anxiety` and `stress` (75% for Prophet). When considering the enclosing week for the hits ($n = 7$), the average hit rates increased to 62% and 66%, for ARIMA and Prophet, respectively. In summary, these results show that individually considering the markers is not enough for accurately forecasting the prevalence of mental health and emotions.

*4.2 Multivariate forecasting*

In VAR, the best results for each dimension did not follow the same trend as in univariate forecasting in which all best results shred the same training period (21 days). Instead, we observed variations in the training days with the best performance in VAR. Two `mental health` dimensions required more training days than the `emotion` ones. For `positive` and `negative emotions`, the lowest MAPE values were achieved when considering 7 training days, with similar results as those observed for ARIMA with 21 days. On the other hand, for `anxiety` and `depression`, only considering 7 training days tripled the errors observed for 14 or 21 days. For `depression`, using 14 training days led to a performance similar to that of ARIMA (with fewer training days than ARIMA). In the case of `stress` using 7 days, the average errors for the markers were higher than those for the other training days, but small statistically significant differences were observed only for two markers (disgust and health). This was caused by a highly skewed error distribution towards the high values, as shown by the standard deviations in Table II. Except for `stress`, in general, errors for VAR had a smaller range than for ARIMA and Prophet. As for `emotions`, ARIMA and Prophet showed more stability in their results, which was evidenced by lower ranges in their forecasting. Minimum errors were also reduced by using VAR up to an 95% (e.g., the *sadness* markers for `anxiety`).

The comparison between the real and forecast series revealed small-to-large differences for all markers, which is consistent with the univariate analysis results. Finally, as regards hit rate (Table II), VAR seemed to improve the results of ARIMA and Prophet for most dimensions. In average, considering a window of $n=3$ led to the discovery of 64% of the peaks. The highest improvements were observed for `depression` (104% regarding ARIMA) and `positive emotions` (42% regarding Prophet). As previously mentioned, despite a few days with high errors, the forecast series with 7 training days for `stress` maintained a shape similar to the original one, with an average hit rate improvement of 16% when compared to univariate analysis (*n*=7). For `positive emotions` hit rates were improved a 22% and 32% in average, for n=2 and n=3, while for `negative emotions` hit rates were improved by a 11% and a 20% in average, for n=2 and n=3, respectively.

The observed error differences for the same marker in the different dimensions (e.g., *disgust* belongs to both `depression` and `stress`) show the effect of the interrelations between markers in the forecast. Overall, multivariate results showed that considering the (implicit) relationships among markers improves the quality of forecasting, while reducing the training days needed.

*4.3 Deep learning forecasting*

The analysis of the MAPE values for RNN showed the best results when using 7 training days, which is the preferred option over larger training periods. In fact, errors were lower or statically similar to errors observed when including more days in the training period. The lack of significant differences means that adding training days does not necessarily improve results. This phenomenon agrees with the observations of Suhara et al. (2017) about psychological surveys, who stated that after 14 training days no improvements were observed, showing that the information of up to the last 14 days is relevant for estimating future psychological states.

When compared to the errors in ARIMA, Prophet and VAR, for the three `mental health` dimensions, we found medium-to-large statistically significant differences favoring the deep learning strategy. In this case, the largest error differences were in *emotional* (66%) for `depression`. Statistically significant differences favoring RNN were found for all markers. This means that the deep learning forecasting improved the best error levels of simpler techniques, while requiring less training data. A practical benefit of the RNN strategy is that enables decision makers to run forecasts and analyses sooner than with the other strategies. In the case of `emotions`, errors were already low for the best performing univariate and multivariate forecasts. Thus, the error variations with deep learning were smaller than those for `mental health`. In this regard, in some cases we did not observe significant differences with ARIMA. This trend was exhibited, for example, in *hate like* and *love* for `negative` and `positive emotions`, respectively.

In several cases, we observed no statistically significant differences between the real and the forecast marker values, which implies that the deep learning strategy can more adequately learn the marker trends. These results, in turn, led to better approximations of the real gradient of the dimensions (as shown in Figure 2), and thus higher hit rates. Table III shows that the deep learning forecasting allowed to achieve the highest hit rates, when considering a three day-window, and it also outperformed almost all univariate and multivariate hit rates. The two exceptions were `stress` and `negative emotions`, for which RNN achieved the same hit rate as VAR. We argue that the better performance of RNN when compared to VAR can be attributed to the ability of neural networks to learn seasonal behaviors and complex (not linear) interactions among the markers.

Finally, we answer our research questions as follows:

> **RQ#1** - Results showed that the training periods achieving the best forecasts depended on the time series strategy. While simpler univariate forecasting required 21 days, for multivariate forecasting 7 days were enough in most cases to improve the best univariate results. These observations could imply that simple strategies require more training days to compensate for disregarding the interactions between the individual time series.

> **RQ#2** - Differences were observed between the forecasts for the mental health and emotions dimensions. Emotions were simpler to forecast, achieving small errors even for the univariate strategies. This fact can be due to the interactions between the emotional markers being not as relevant (for prediction) as those for the mental health markers. Despite the multivariate improvements (over the univariate strategies) were lower than those for the mental health markers, multivariate techniques still worked well with a training period of 7 days.

> **RQ#3** - The deep learning strategy achieved the best forecasts (in terms of quality), as they were able to leverage on the interactions among the markers, and thus to represent better the original trends (and peaks) of the dimensions. Moreover, using deep learning minimized the training period needed. Therefore, the emergence of high prevalence periods of emotions and mental health disorders can be anticipated even when limited data is available.

**5. Conclusions**

Crises, like the COVID-19 pandemic, affect society from multiple points of view, not only limited to physical health, but also to mental health, economics and politics. Hence, it is crucial to understand how people react to different events and how their emotions and mental health evolve over time as a crisis develops. Several studies have focused either on individual users or relied on surveys or data explicitly provided by users, which limits the scope of the study and the generalization of the analysis. We believe that social media allows analyzing the mental health of a given population (or group of individuals) at a larger scale and for longer time periods.

From a practical point of view, our approach provides a tool for forecasting mental health markers and estimating peaks with a high prevalence of (possible) mental health disorders in short- to medium-term time horizons. We showed how different time series strategies offer different prediction capabilities for the task. In our experiments with a dataset of COVID-19-related tweets for Argentina, the forecasting based on neural networks provided the best results with a short training period. In average, the neural network strategy detected 80% of the high prevalence peaks, with an average improvement of 66% regarding the univariate strategy. Differences were observed between the forecasts the mental health and emotion dimensions, presumably due to differences in the interactions between the markers. In this sense, the deep learning forecasts better captured those marker interactions and thus the original trends of the time series. Although the empirical work can benefit from a larger validation study, we believe it shows evidence that the proposed approach can be a valuable asset for the design and monitoring of health prevention and communication policies (Holmes et al., 2020). The framework behind our approach is general, and it can apply to other social media or psycho-social theories.

We envision a number of aspects to be tackled as future works. First, this study focused so far on the society as a whole, without considering demographic differences among Twitter users. It would be interesting to classify users according to demographic characteristics (or other segmentation criteria) in order to observe how groups of users express during a crisis and how their mental health or emotions vary over time. Second, based on the study of the COVID-19 pandemic in Argentina, we could compare how the pandemic manifested in neighbor countries and how the different policies adopted by each government (in combination with cultural dimensions and political orientation) can affect the manifestations of mental health disorders. Third, as Twitter is not the only site in which the government has an official presence, we could account for the perspectives provided by the users of the different sites, and then determine whether mental health manifest similarly across sites. At last, we would like to support decision makers by identifying which time events in the training period (of the series) were more influential for the forecast markers as well as for the dimension peaks. An interesting alternative for this identification are the recent developments on explainability for Machine Learning models.

**Notes**

[1] Available at: https://data.mendeley.com/datasets/nv8k69y59d/2
[2] Available at: https://github.com/knife982000/FakingIt
[3] https://fasttext.cc/
[4] https://www.ibm.com/watson/services/language-translator/
[5] Available at: http://bit.ly/3pFH8GS After review, this will be moved to a public repository.